%% file: Paper.tex
\documentclass[aps, 11pt, final, notitlepage, oneside, twocolumn, nobibnotes, nofootinbib,
 superscriptaddress,  noshowpacs,  centertags, tightenlines]
{revtex4}

\input{sao_cmd_author.tex}

\begin{document}
\selectlanguage{english}

\keywords{stars: fundamental parameters -- stars: variables: S Doradus -- stars: mass-loss -- stars: winds, outflows -- X-rays: binaries}

%

\title{Modeling of extended atmospheres with temperatures below 40,000\,K}

\author{\firstname{A.}~\surname{Kostenkov}}
 \email{kostenkov@sao.ru}
 \affiliation{\saoname}
 \affiliation{Saint Petersburg State University, 7/9 Universitetskaya Emb., 199034, Saint Petersburg, Russia}

\author{\firstname{A.}~\surname{Vinokurov}}
 \affiliation{\saoname}

\author{\firstname{Y.}~\surname{Solovyeva}}
 \affiliation{\saoname}

\author{\firstname{K.}~\surname{Atapin}}
 \affiliation{Sternberg Astronomical Institute, Moscow State University, Universitetsky pr., 13, Moscow, 119991, Russia}

\author{\firstname{S.}~\surname{Fabrika}}
 \affiliation{\saoname}
 \affiliation{Kazan Federal University, Kremlevskaya 18, 420008 Kazan, Russia}

\begin{abstract}
We calculate models of extended stellar atmospheres with a temperature in the range of $12000 \text{-} 40000$ K and a mass loss rate of $10^{-6} \text{-} 10^{-4} M_{\odot} \text{yr}^{-1}$. A large number of objects with emission spectra, such as luminous blue variables (LBV), Fe\,II - emission line stars, Of/late-WN stars, and even ultraluminous X-ray sources (ULXs) often have effective temperatures in this range. The paper presents the results of model grids calculating in the form of equivalent width diagrams for the selected lines of hydrogen, He, Si, and Fe, as well as the results of studies of some emission objects using the calculated models.
\end{abstract}

\maketitle

\section{Introduction}           
\label{sect:intro}

Strong radiative-driven winds are characteristic of most classes of massive stars. The most known among them are luminous blue variables (LBVs), Fe\,II - emission line stars, warm hypergiants, Of/late-WN stars, hot and intermediate supergiants, B[e] supergiants \cite{Humphreys2014}. B[e] stars differ from the others in that their outflows are concentrated near the equatorial plane, while the winds of stars of the other classes are much closer to spherical symmetry. A study of the winds is the key to understanding the physics of these stars.

A precise determination of fundamental stellar ($L$, $T_{\text{eff}}$\footnote{$T_{\text{eff}}$ is the temperature at Rosseland optical depth $\tau =2/3$}, $R_{*}$) and wind ($\dot {M}$, $V_{\infty}$, $\beta$) parameters from optical spectra is a difficult task because conditions in the wind are far from equilibrium. Reliable results for a particular object could be reached only by simulations of its spectrum based on the physical model of its extended atmosphere. The most advanced codes developed for this purpose are CMFGEN \cite{Hillier1998} and PoWR \cite{Hamann2006}. Both of them are widely used to simulate spectra of LBV and Wolf-Rayet (WR) stars.

Not only stars exhibit stellar-like winds. Spectra with strong and broad emission lines produced by outflowing matter are also observed in ultraluminous X-ray sources (ULXs). ULXs are bright, point-like, off-nucleus, extragalactic sources with X-ray luminosities above $10^{39}$ erg\,s$^{-1}$. Studies of X-ray and optical spectra of ULXs on the last decade have revealed that most of these objects are stellar mass black holes or even neutron stars accreting in a super-Eddington regime (\cite{Kaaret2017} and references therein). \cite{Fabrika2015} have shown that optical spectra of ULXs are similar to those of LBV and WNLh stars (late nitrogen-sequence WR stars with hydrogen lines). Analyzing the ratio between the widths of the H$\alpha$ and He\,II $\lambda$4686 lines in the spectra of ULXs, LBVs, WNLs and Galactic stellar-mass black hole X-ray binaries in outbursts, they concluded that emission spectra of ULXs are formed in the winds of supercritical disks (which have to be similar to stellar winds) rather than in self-irradiated accretion disks. Rapid variability of the emission lines itself and their radial velocities made it possible to rule out with a high probability the donor stars as a possible origin of observed emission spectra. We believe that the similarity between the ULX outflows and stellar winds allows to use, at least as a first approximation, spherically symmetric atmosphere models to estimate parameters of the ULX winds.

An accurate fitting of the synthetic spectra to the observed ones requires to calculate many models and spend a lot of computational time. This motivated us to compute the grid of models that could provide a `quick-look' at the wind parameters. In this manuscript we present our first results on grid calculations. Here we publish two grids of models, one for LBV-like stars and one for ULXs (for the certain chemical abundance), which cover parameters  typical for objects of these classes. Our grids allow to estimate the photosphere temperature and the wind mass-loss rate using equivalent widths (EW) of certain lines measured from the observed spectrum. Also we have tested our results on the sample of sources (LBVs: WS-1, AG\,Car, P\,Cyg, Var\,15, Var\,A-1, the Galacic LBV candidate (cLBV) MN112; Of/late-WN stars: M31-004242.33, M31-004334.50, M31-004341.84; ULXs: Holmerg\,II~X-1, NGC\,4559~X-7, NGC\,5204~X-1, UGC\,6456~ULX).

\section{Methods}
\label{sect:method}

\begin{table*}
\setcaptionmargin{0mm} \onelinecaptionstrue \captionstyle{normal}
 \caption{Fundamental parametres for LBV and ULX model grids.}
 \label{param_table} 
 \medskip
\begin{tabular}{c|c|c}
 \hline
   & LBV grid  & ULX grid  \\
 \hline
  Temperature (at $\tau \approx20$) $[\text{kK}]$ & $4.08 \leqslant \log{T_{*}} \leqslant 4.60$ & $4.08 \leqslant \log{T_{*}} \leqslant 4.60$ \\
  & Step $0.025$ & Step $0.025$ \\
  \hline
  Mass-loss rate $[M_{\odot}\text{yr}^{-1}]$ & $-5.30 \leqslant \log{\dot{M}} \leqslant -3.90$ & $-6.35 \leqslant \log{\dot{M}} \leqslant -4.50$ \\
  & Step 0.10 & Step 0.143\\
  \hline
  Luminosity $[L_{\odot}]$ & $\log{L_{*}}=5.3$ & $\log{L_{*}}=5.0$\\
  Terminal velocity $[\text{km s}^{-1}]$ & 300 & 300\\
  Photospheric velocity $[\text{km s}^{-1}]$ & 30 & 30\\
  Turbulent velocity $[\text{km\,s}^{-1}]$ & 15 & 15\\
  Wind velocity $\beta$-law & 1.0 & 1.0 \\ 
  Volume filling-factor $f$ & 0.3 & 0.3\\ 
  $X_{\text{H}}$ $[\%]$ & 39.7 & 70.1\\
  $X_\text{N}/ X_{\odot}$ & 10.8 & 0.2 \\
  $X_\text{C}/ X_{\odot}$ & 0.4 & 0.2 \\
  $X_\text{Si}/ X_{\odot}$ & 1.0 & 0.2\\
  $X_\text{Fe}/ X_{\odot}$ & 1.0 & 0.2\\
 \hline
\end{tabular}
\end{table*}

To compute the models we used the iterative non-LTE line-blanketing code CMFGEN \cite{Hillier1998}. Models of this code are determined by fundamental stellar parameters ($L$, $T_{\text{eff}}$, $R_{*}$), wind properties ($\dot{M}$, $V_{\infty}$, $\beta$), chemical abundances and some auxiliary parameters like the turbulent velocity $V_{\text{turb}}$ and filling-factor $f$. The velocity law is described by the equation \cite{Hillier1989} that combine the isothermal atmosphere with effective scale height $h$ and the $\beta$ law in the wind \cite{Lamers1996}.

The model parameters used in our grids are listed in Tab.\,\ref{param_table}. For the LBV grid, we were guided by the fundamental parameters and chemical abundances obtained by several authors from spectral modeling \cite{Najarro2, Groh2009, Maryeva2010, Mahy2016}. We included the following ions: H (I), He (I, II), C (I, II, III), N (I, II, III, IV), Si (II, III, IV), Fe (II, III, IV, V, VI, VII), whose lines are seen in spectra of the objects of our interest (e.g. \cite{Humphreys2014, Humphreys2016_II}). Levels of the neutral N and C are needed for a more accurate calculation of the wind structure far from the star surface. In the case of ULXs, the ranges of the temperature and mass loss rate were chosen based on our preliminary modeling of the spectra of some small sample of ULXs. The hydrogen abundance was assumed to be solar because this is favored by the observed He\,II/He\,I emission line ratios \cite{Fabrika2015}. Metallicity of ULXs is usually considered to be equal to the metallicity of the parent galaxy, and in most cases it is sub-solar (e.g., \cite{Kaaret2017}). We assumed metallicity equaled 0.2 of the solar value (Tab.\,\ref{param_table}).

We represent the grids as equivalent width diagrams of the selected lines for models with different mass-loss rates and temperatures. The lines were selected to be the most sensitive to the model parameters in the particular parameter range. Actually, some of the fundamental stellar parameters can be determined almost independently even from a relatively small optical spectral range. The temperature can be estimated from the EW of the lines that correspond to different ionization stages of the same chemical element. For example, the temperature can be determined using the EW of the He\,II $\lambda$4686 line and resonance He\,I lines. However, this method is not applicable to objects with a temperature less than $\approx25000$\,K. Most LBVs have effective temperatures less than $25000\,$K, while ULXs show hotter spectra with a strong He\,II $\lambda$4686 line. Bearing this in mind, we calculated the models in the temperature range 12000--40000\,K, which covers all the objects studied in this work.

To estimate the temperature within the range of $12000 \text{--} 25000$\,K, one can use the lines of Fe\,II, Si\,II and the hydrogen Balmer series lines. A relatively large abundance of the ions Fe\,II and Si\,II is caused by the charge-exchange reactions that transfer electrons from neutral hydrogen to Fe\,III and Si\,III. Since the proportion between the neutral and ionized hydrogen depends on both the temperature and the mass-loss rate (fraction of the neutral hydrogen increases with the wind density), Fe\,II and Si\,II lines are also quite sensitive to these parameters. We have chosen the Fe\,II $\lambda$5169 and Si\,II $\lambda$6371 lines. These lines are mostly formed in optically thin outer parts of the wind, and their profiles have negligible absorption components. It decreases potential errors of the EW measuremnts which may occur due to isufficient resolution of the spectra (see Sec.\,\ref{sect:res}).

Spectra of the majority of LBV-like star are dominated by hydrogen and helium lines.
There are many singlet (e.g. He\,I $\lambda$5015, $\lambda$6678) and triplet (e.g. He\,I $\lambda$4471, $\lambda$5876, $\lambda$7065) helium lines in the optical region. The strength of the singlet lines significantly depends on the velocity structure of the wind. Lower velocities near the sonic point correspond to higher densities in this region. It makes the choice of the velocity law $\beta$ and the particular velocity $V_0$ at the sonic point (generally referred to as the photospheric velocity) very important. However, the velocity structure near the sonic point cannot be determined reliably from the optical range; one needs to analyse the IR spectra. As a result, most He\,I lines must not be used for the temperature estimates. Only the He\,I $\lambda$5876 and He\,II $\lambda$4686 lines can be involved, because the EW of the triplet He\,I $\lambda$5876 line is not significantly affected by velocity structure close to the star surface. 

Additionally, the strength of the lines that are formed in dip regions of the wind, where the Thomson scattering opacity becomes very high, is strongly affected by the turbulent velocity $V_{\text{turb}}$. Since the velocities of the bulk and turbulent motions of the wind matter are comparable in these regions, there is a high probability that the photon from the line wing (or even from continuum) will be redistributed into the line core. After such a redistribution, the photon can be absorbed and re-emitted in the line, thus enhance its strength \cite{Hillier1991}. In particular, this mechanism significantly increases the equivalent width of the He\,II $\lambda$4686 line. Also, the lines of high optical depth that are formed in outer parts of the extended atmosphere, where the wind velocity is high and the Thomson opacity is negligible, may be asymmetrically shifted toward red parts of the spectrum because the photons scattered to the red wing of the line cannot be reabsorbed and leave the medium while the photons of the blue wing can be absorbed once again \cite{Hillier1991}. A significant difference in temperature estimates obtained from several diagrams may indicate a large contribution of scattering to strength of some lines (see Sec.\,\ref{sect:res}, star WS-1). 

We have chosen $V_0=30\text{ km\,s}^{-1}$ and $V_{\text{turb}}=15\text{ km\,s}^{-1}$ for our model. Similar values of the photospheric and turbulent velocities were used by other authors in their modeling of the spectra of several LBVs \cite{Najarro1997_IR, Groh2009}.

The mass-loss rate can be obtained from equivalent widths of Balmer series lines in conjunction with the temperature estimates derived from the methods described above. For the objects with temperature $T \lesssim 20000\,$K, the strength of absorption components of Balmer lines is determined by the ionization state of the wind, and even slight variations of the temperature and mass-loss rate can significantly change the ionization structure of the wind \cite{Najarro2}. The ratio between the absorption and emission components of the Balmer lines could be used for the temperature and mass-loss rate estimates, but very carefully, with detailed modeling, because these lines have strong electron-scattering wings which makes it difficult to distinguished their absorption and emission components in low resolution spectra. 
Applying equivalent width diagrams in this case discussed in Sec.\,\ref{sect:res}.

To introduce clumping, the CMFGEN code uses a simple filling-factor approach \cite{Hillier1999}. The emissivities and opacities scale as the square of the density for thermal processes, while the electron scattering scale linearly. This reduce the relative strength of the electron-scattering wings in strong emission lines compared to the smooth-wind model \cite{Hillier1991}. The filling factor $f$ and the mass-loss rate are bound parameters. The mass-loss rates of the smooth-wind model $\dot{M}_0$ and that of the model with clumping $\dot{M}_{\text{cl}}$ describing the same observed spectrum are related as $\dot{M}_{\text{cl}}= \dot{M} f^{-0.5}$. We assumed $f=0.3$ for our models. To turn to another filling-factor, one can use the following relation $\dot{M}_{\text{new}}=\dot{M}_{\text{old}}\sqrt{f_{\text{new}}/0.3}$

The terminal wind velocity $V_{\infty}$ can be determined independently of other parameters by the FWHM measurements of forbidden lines which are formed in outer parts of the wind expanding with near constant velocity. For example, the terminal wind velocity of LBV-like stars with temperature $T \lesssim 20000$\,K can be obtained from several [Fe\,II] lines or the [N\,II] $\lambda$5755 line \cite{Stahl1991}. If such lines are absent in the spectrum, the terminal wind velocity can be estimated, after spectral resolution correction, by the blue shift of the absorption component respect to the emission peak in strong hydrogen and He\,I lines (e.g. \cite{Maryeva2010}). However, results obtained with this method may significantly depend on $\beta$ in the wind velocity law.

Our models were calculated assuming one certain luminosity (Tab\,\ref{param_table}). It is possible to rescale the result to new (observed) luminosity bearing in mind that the effective temperature and the wind density parameter $\dot{M} / (V_{\infty}\, R_{*}^{3/2})$ have to be conserved to keep the equivalent widths of hydrogen and helium lines \cite{Schmutz1988}. As a first approximation, one can use  $\dot{M} \sim R_{*}^{3/2}$ and $L\sim R_{*}^{2}$, which yields the following relation: 

\begin{equation}
\begin{array}{r}
    \dot{M}_{\text{new}}=\dot{M}_{\text{old}}\sqrt{f_{\text{new}}/0.3} \\ (V_{\text{new}}/V_{\text{old}}) \cdot (L_{\text{new}}/L_{\text{old}})^{3/4}
\end{array}    
    \label{eq1}
\end{equation}

Nevertheless, changes of the mass-loss rate (and thus the wind density) also change a position of the bottom boundary of the atmosphere and affect the effective temperature, so the above equation is not quite accurate. More precise relations between the LBV model parameters has been proposed by \cite{Najarro}.

\begin{figure*}
 \setcaptionmargin{5mm} \onelinecaptionstrue \captionstyle{normal}
  \includegraphics[scale=0.15]{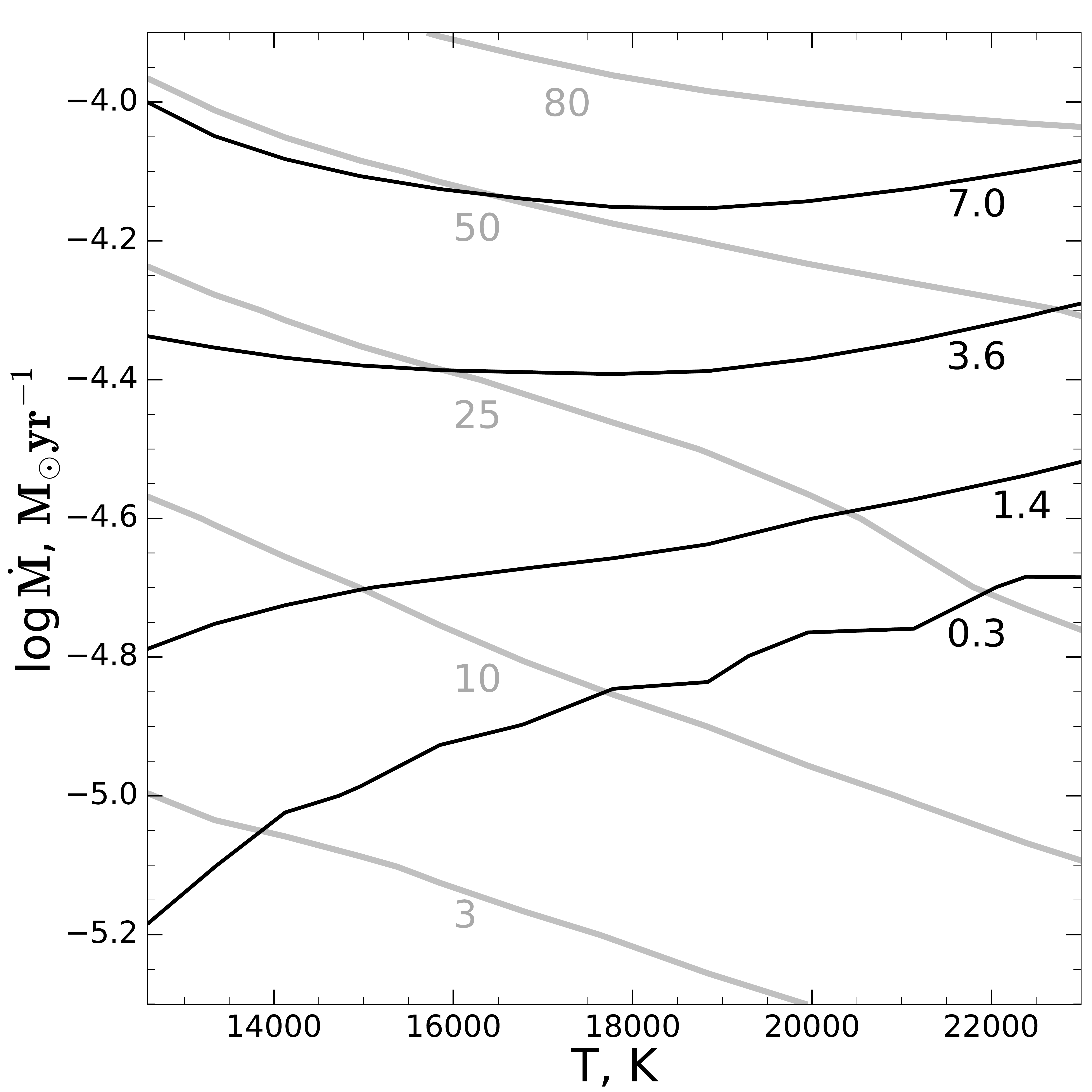}
  \includegraphics[scale=0.15]{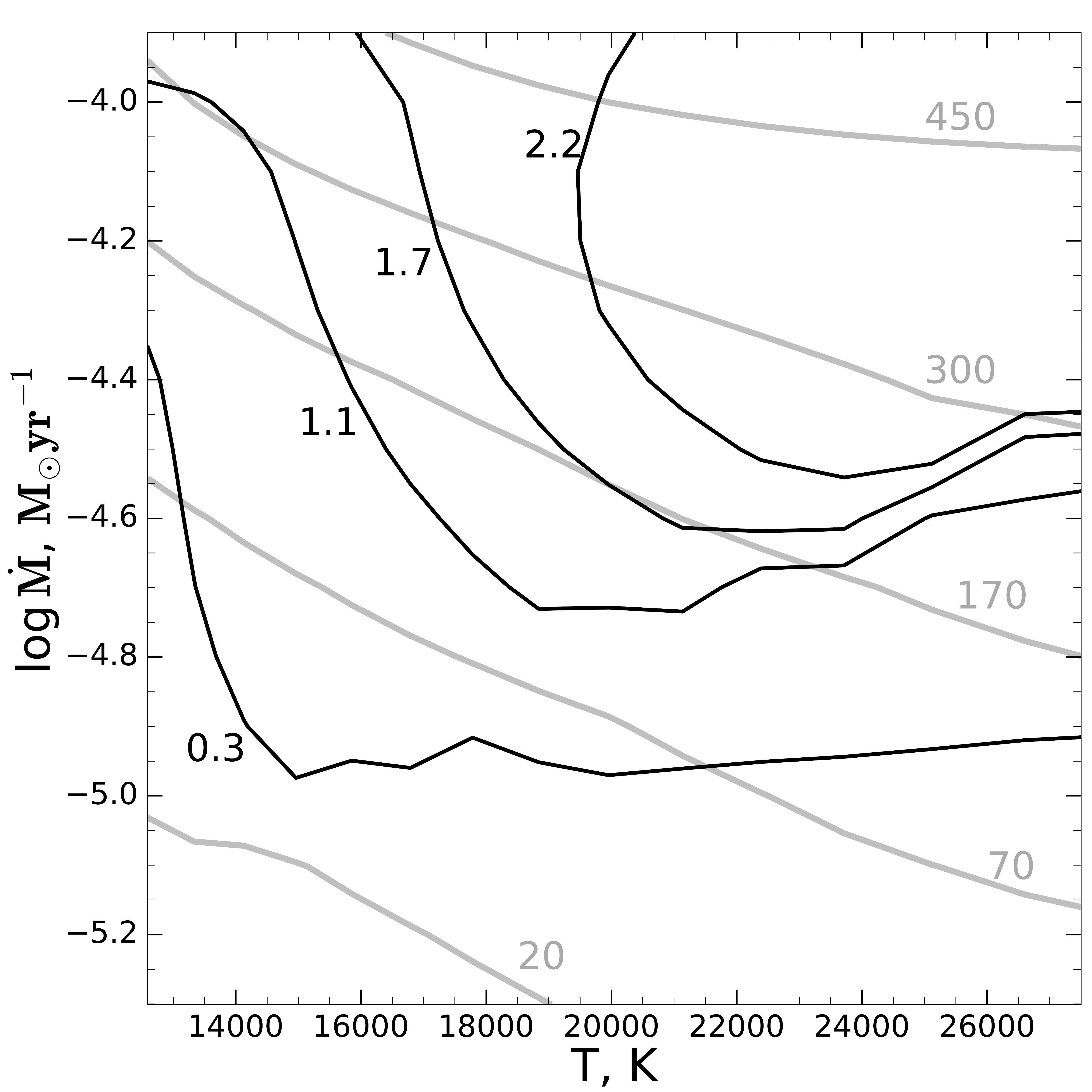}
  \includegraphics[scale=0.15]{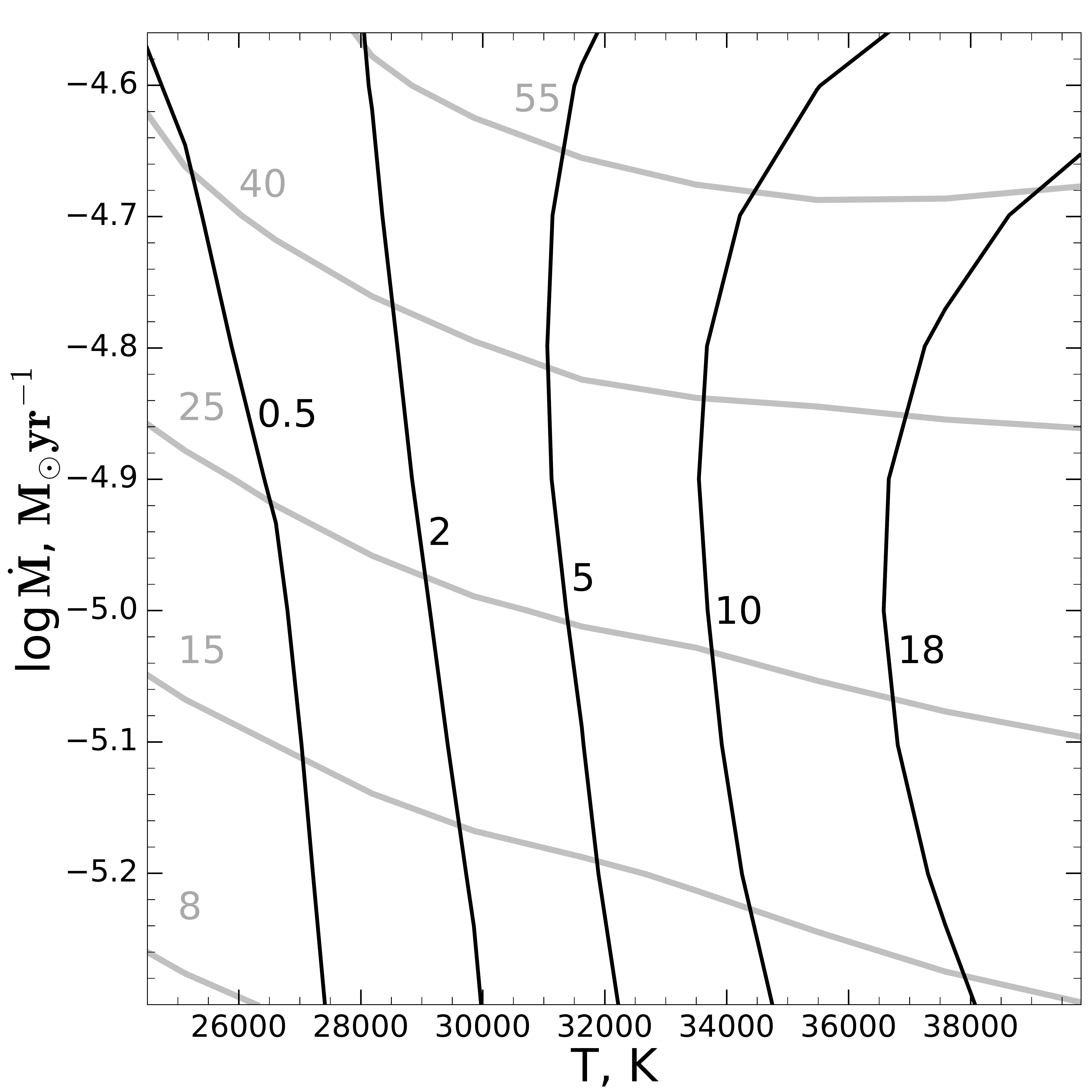}
  \caption{Equivalent widths diagrams of model grids for LBV-like stars (grid parameters: luminosity $\log{L_{*}/L_{\odot}}=5.3$, terminal velocity $V_{\infty}=300\ \text{km s}^{-1}$, volume filling-factor $f=0.3$, and solar metallicity) for case of high resolution; top left: H$\beta$ (grey lines) and Fe\,II $\lambda$5169 (black lines); top right: H$\alpha$ (grey lines) and Si\,II $\lambda$6371 (black lines); bottom: He\,I $\lambda$5876 (grey lines) and He\,II $\lambda$4686 (black lines).}
\label{fig:ew_contour_LBVhigh}
\end{figure*}

\begin{figure*}
 \setcaptionmargin{5mm} \onelinecaptionstrue \captionstyle{normal}
  \includegraphics[scale=0.15]{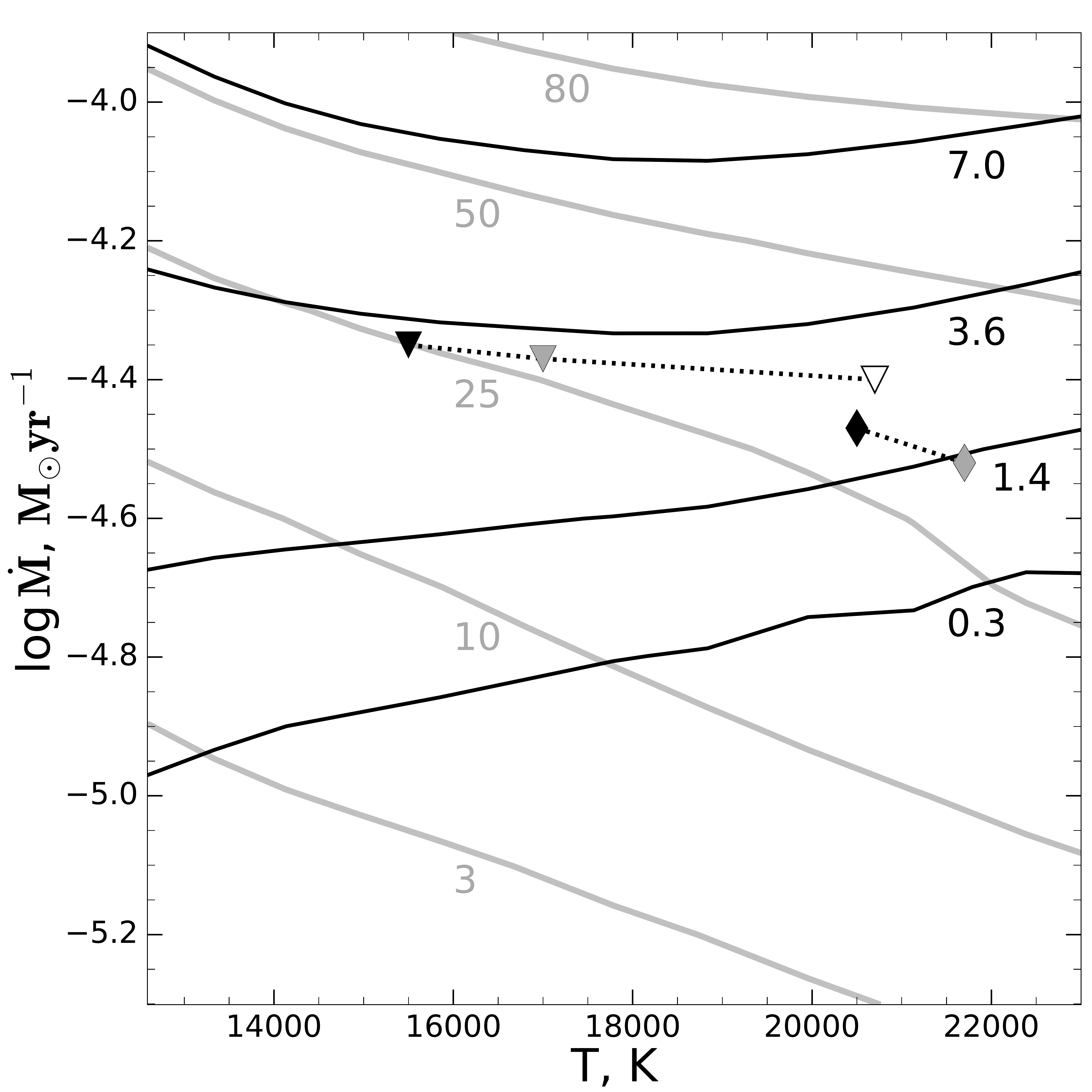}
  \includegraphics[scale=0.15]{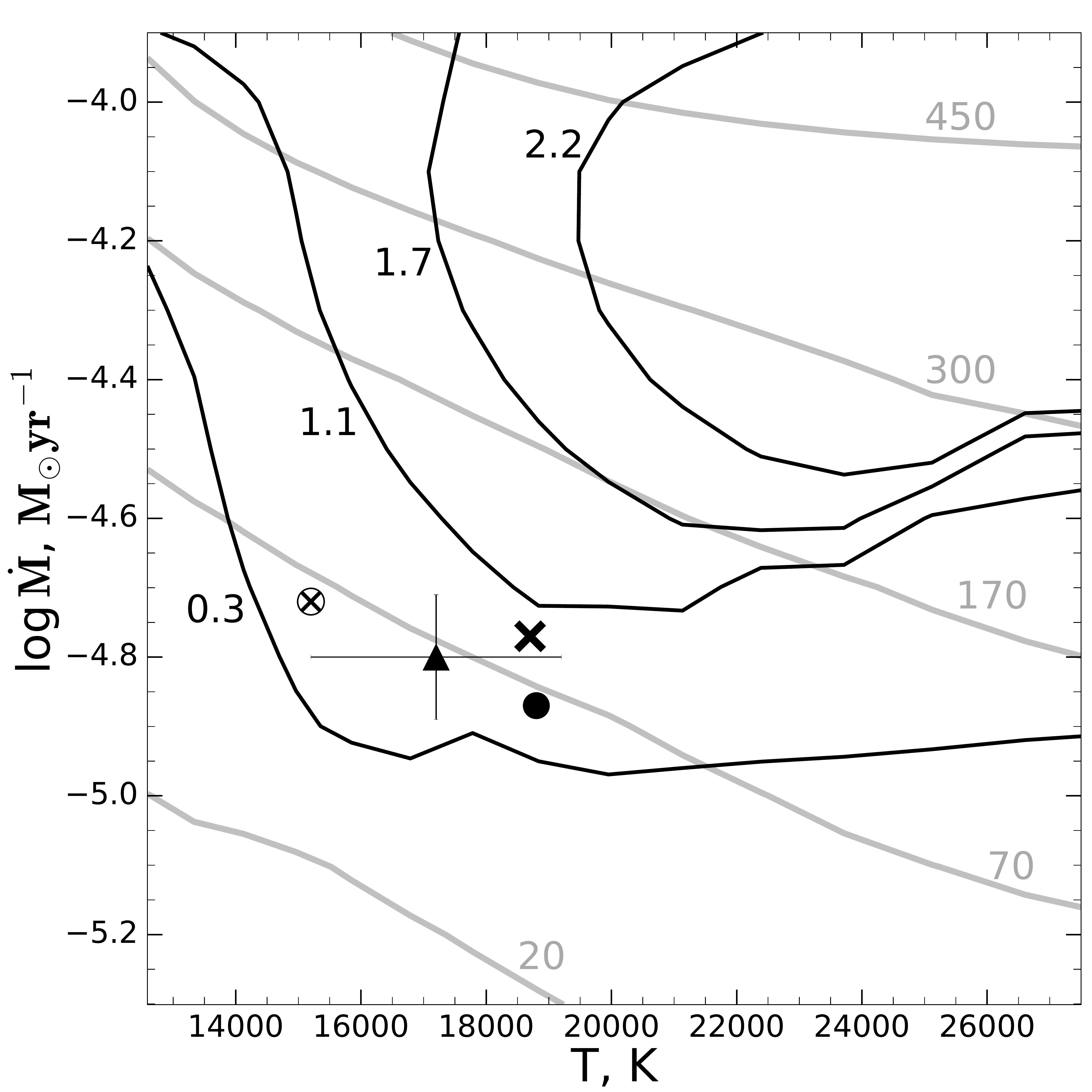}
  \includegraphics[scale=0.15]{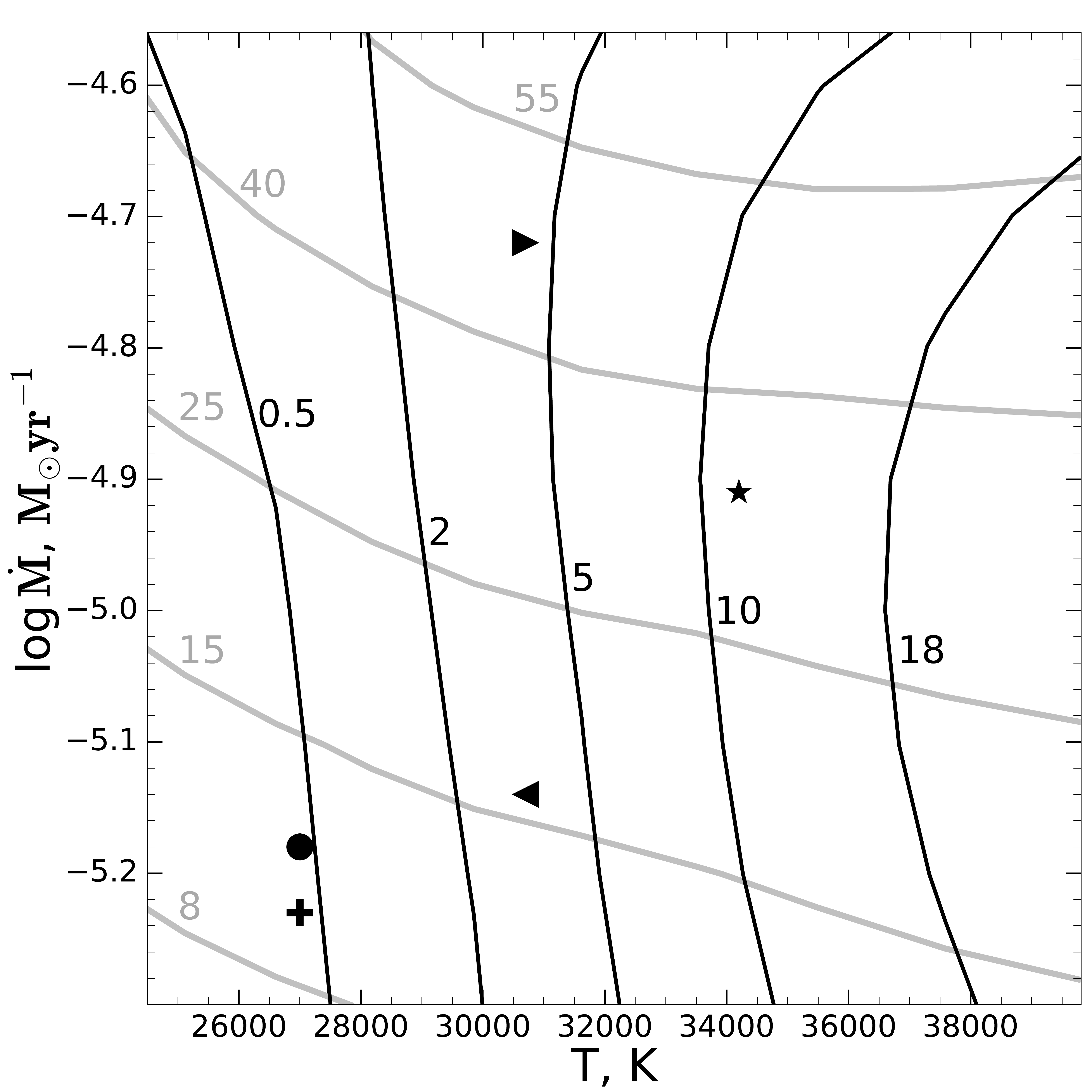}
  \caption{Equivalent widths diagrams are the same as in the Fig.\ref{fig:ew_contour_LBVhigh}, but for spectral resolution of 5 \AA. The following stars are marked with symbols: WS-1 (circles), AG Car (plus), P Cygni (cross), MN112 (up triangle), Var\,A-1 (down triangles), Var\,15 (diamonds), M31-004242.33 (right triangle), M31-004334.50 (star), M31-004341.84 (left triangle). Transitions between different states of Var\,A-1 and Var\,15 are designated by identical symbols with different colors (black, gray and white). The alternative position of WS-1 on the H$\alpha$--Si\,II $\lambda$6371 diagram due to an ambiguous solution is shown by an empty circle with cross (see text for more details).}
\label{fig:ew_contour_LBVlow}
\end{figure*}

\begin{figure*}
 \setcaptionmargin{5mm} \onelinecaptionstrue \captionstyle{normal}
  \includegraphics[scale=0.15]{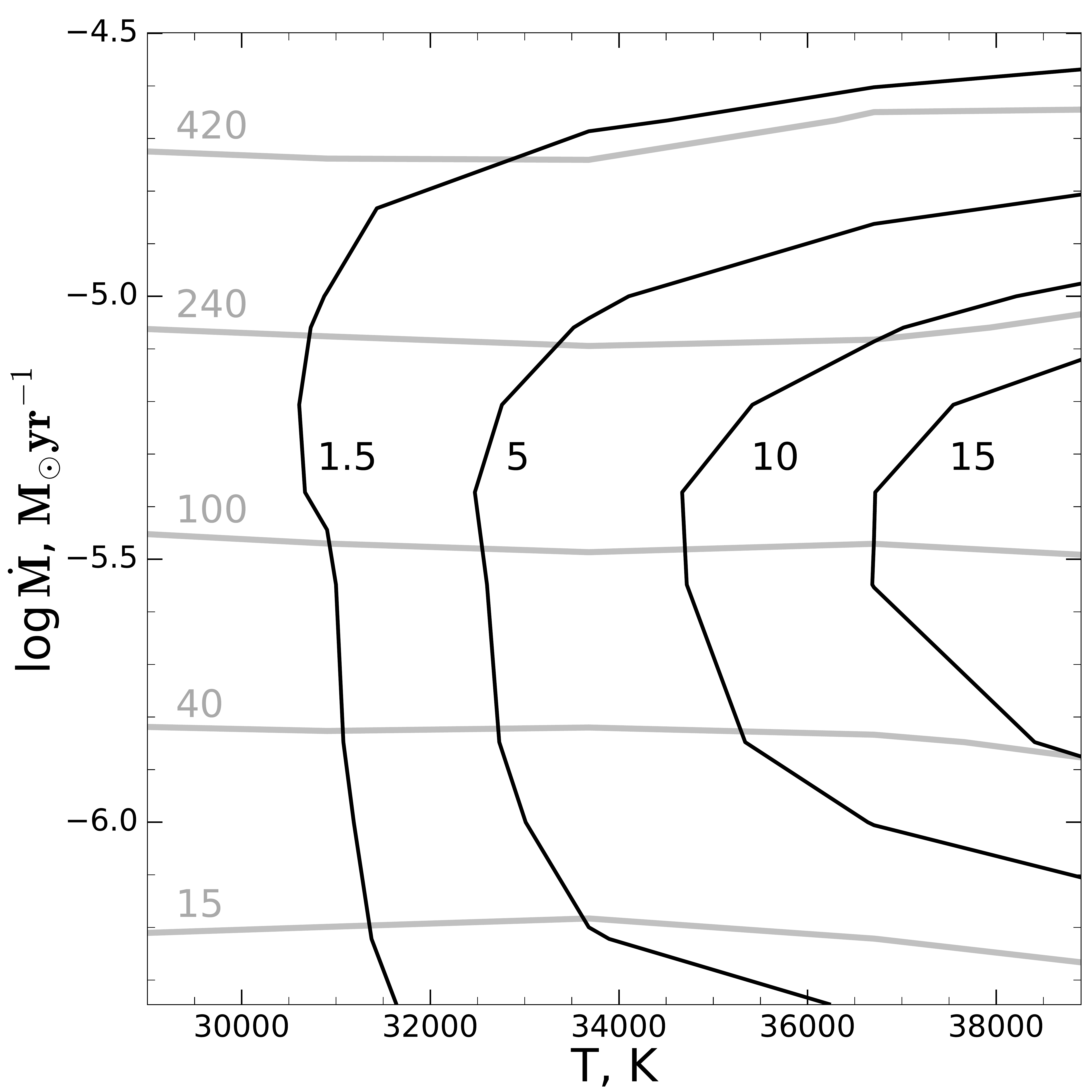}
  \includegraphics[scale=0.15]{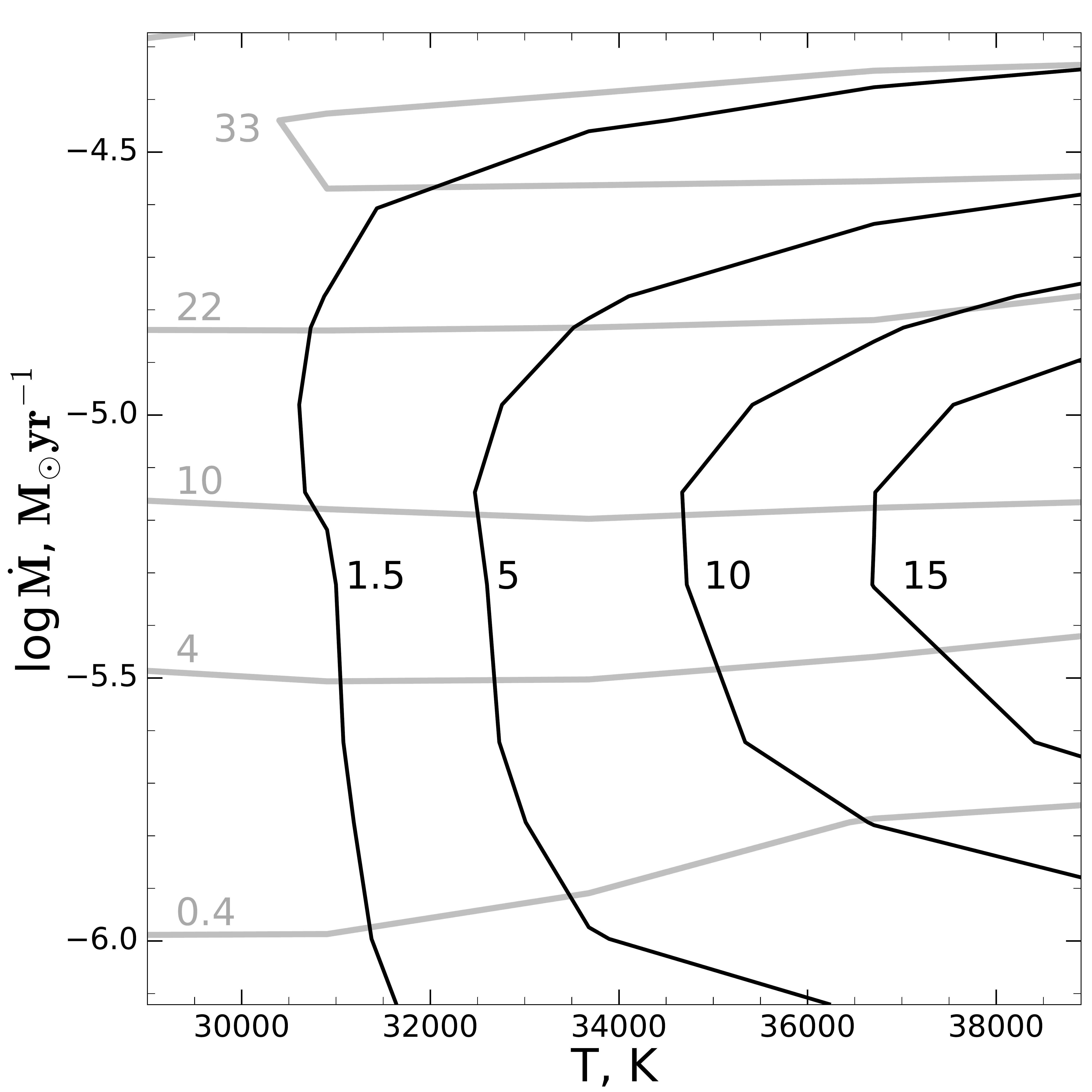}
  \caption{Equivalent widths diagrams of model grids for ULXs (grid parameters: luminosity $\log{L_{*}/L_{\odot}}=5.0$, terminal velocity $V_{\infty}=300\ \text{km s}^{-1}$, volume filling-factor $f=0.3$, and metallicity equaled 0.2 of the solar value) for case of high resolution; left: H$\alpha$ (grey lines) and He\,II $\lambda$4686 (black lines); right: He\,I $\lambda$5876 (grey lines) and He\,II $\lambda$4686 (black lines).}
\label{fig:ew_contour_ULXhigh}
\end{figure*}

\begin{figure*}
 \setcaptionmargin{5mm} \onelinecaptionstrue \captionstyle{normal}
  \includegraphics[scale=0.15]{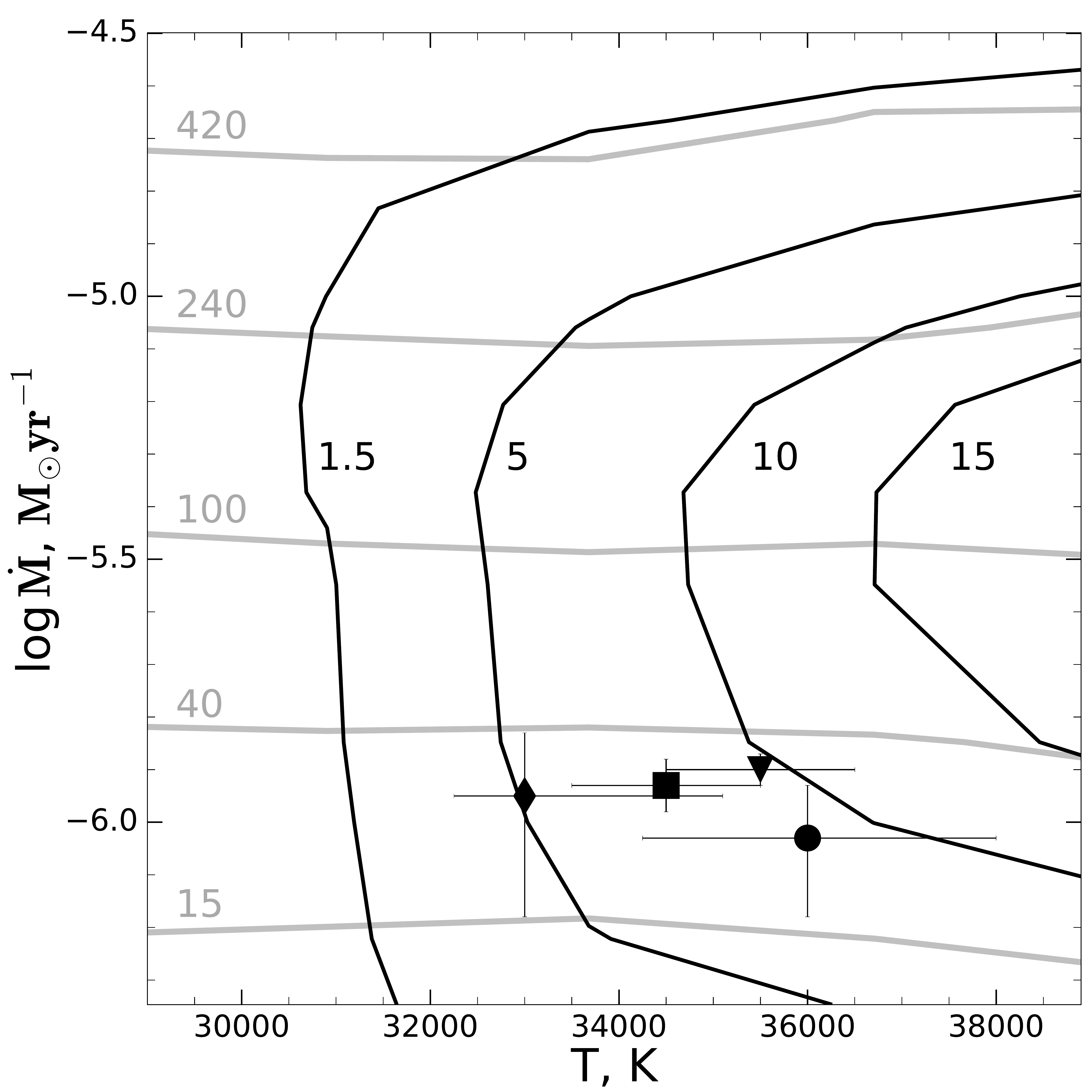}
  \includegraphics[scale=0.15]{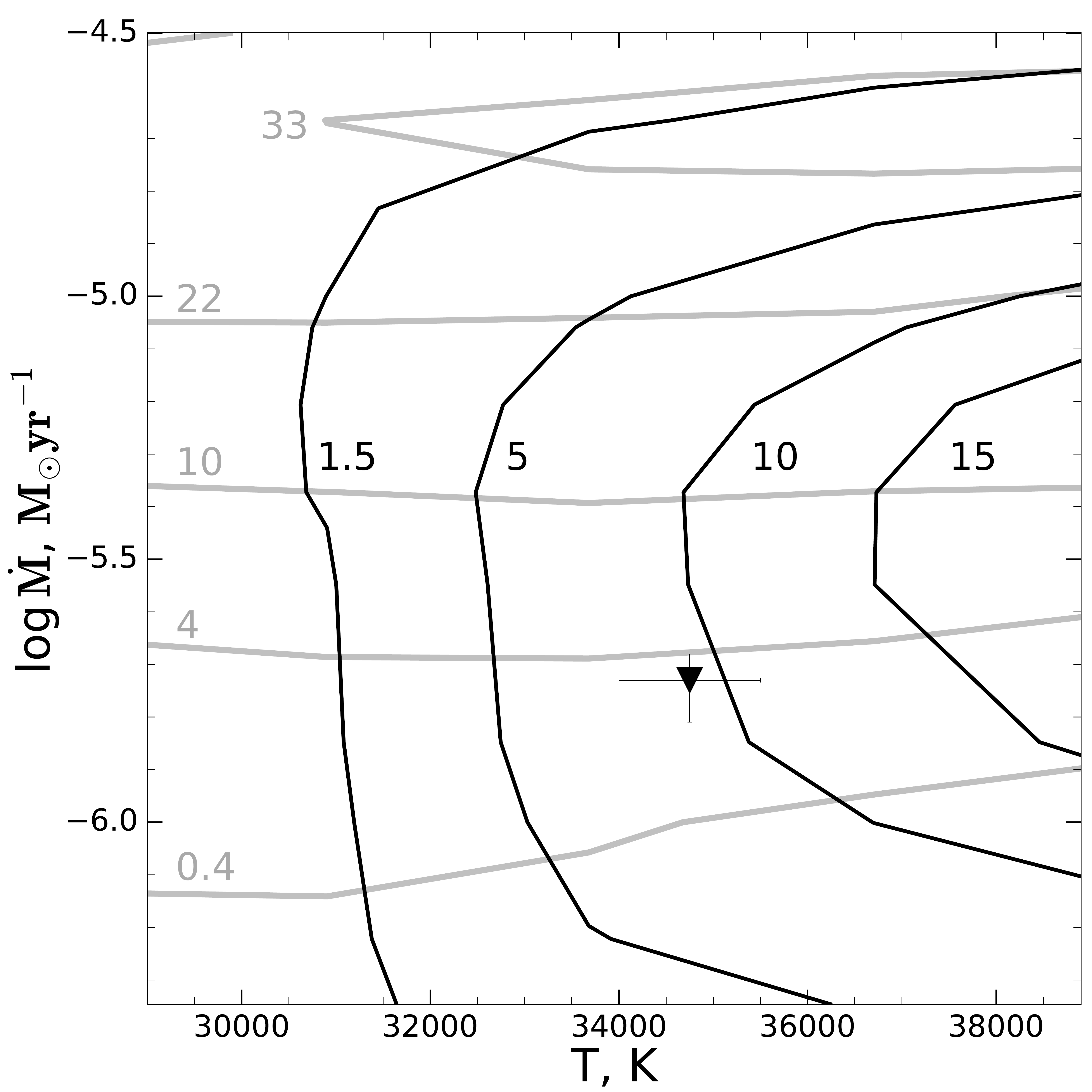}
  \caption{Equivalent widths diagrams are the same as in the Fig.\ref{fig:ew_contour_ULXhigh}, but for spectral resolution of 5\,\AA. The circle marks the position of Holmberg\,II~X-1, the square -- NGC\,5204~X-1, the triangle -- NGC\,4559~X-7, the diamond -- UGC\,6456~ULX.}
\label{fig:ew_contour_ULXlow}
\end{figure*}

\begin{figure*}
 \setcaptionmargin{5mm} \onelinecaptionstrue \captionstyle{normal}
  \includegraphics[scale=0.22]{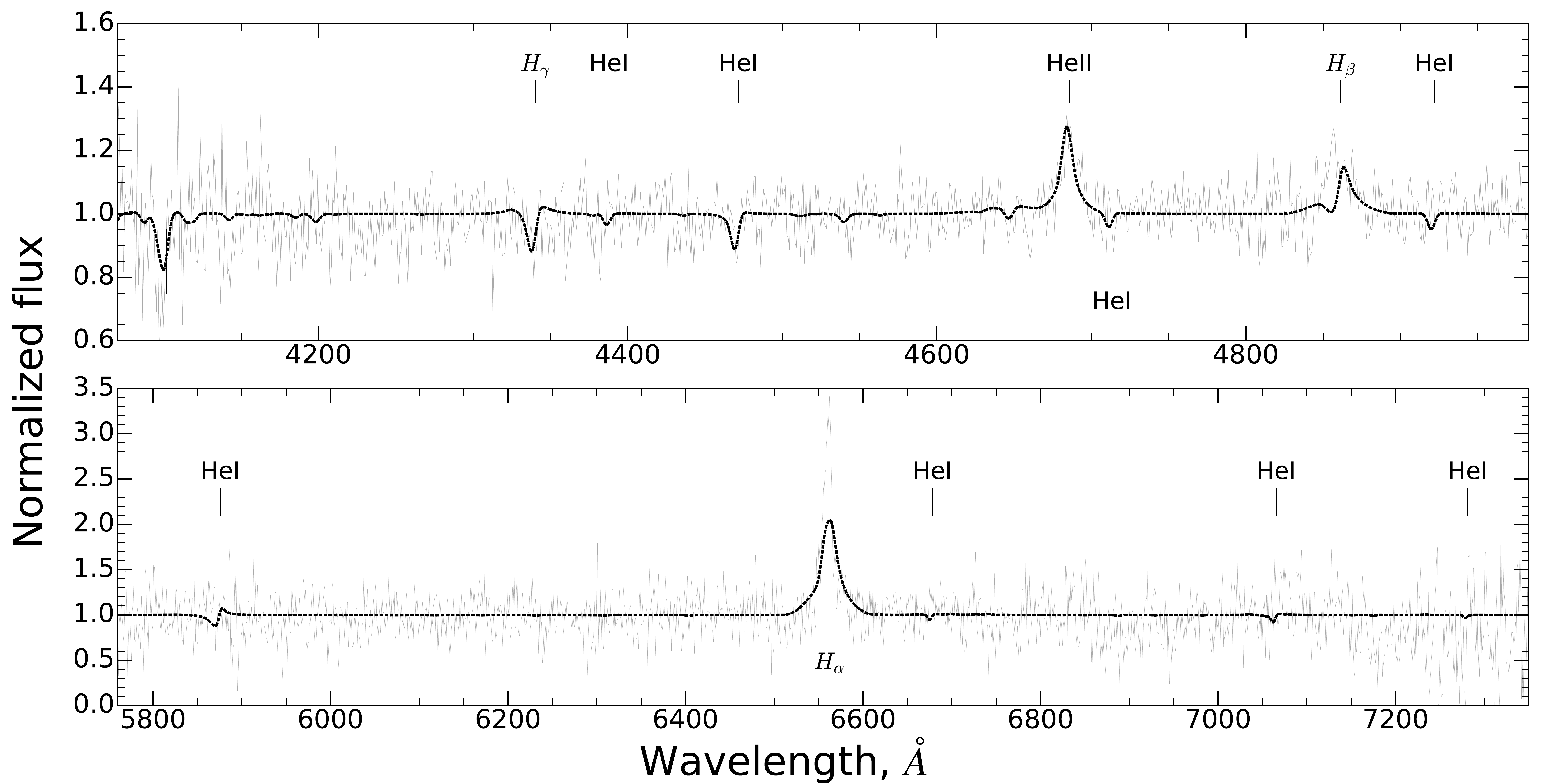}
  \caption{The observed spectrum of ULX in UGC\,6456 (grey solid line) compared with the model spectrum smoothed with a spectral resolution of 5\AA {} (black dashed line).}
\label{fig:ugc}
\end{figure*}

\section{Data}       
\label{sect:obs}
To test our grid we selected four ULXs and ten LBVs, cLBVs and Of/late-WN stars in our Galaxy and in M\,31. The line equivalent widths of three well studied ULXs (Holmberg\,II~X-1, NGC\,5204~X-1 and NGC\,4559~X-7) were taken from \cite{Fabrika2015}. The data for Galactic LBVs AG Car, P Cygni, WS-1 and cLBV MN112 were found in papers \cite{Groh2009}, \cite{Gvaramadze2010}, \cite{Gvaramadze2012} and \cite{Gvaramadze2010}, respectively. Spectra of LBVs (Var\,15, Var\,A-1) and Of/late-WN stars (M31-004242.33, M31-004334.50, M31-004341.84) were taken from the site of Eta Car Research Group\footnote{http://etacar.umn.edu/} (see also \cite{Humphreys2016_II}). For the ULX in the UGC\,6456 galaxy, we additionally carried out detail modeling of its optical spectra and compared obtained result with the parameters estimated  from the grids. UGC\,6456 ULX was observed with the 6-m Russian telescope BTA on 2015 September 7 using the SCORPIO optical reducer \cite{Afanasiev2005} within the range of 4000-5700 \AA{} and 5700-7500 \AA{}, the resolution was 5.3 \AA{}. Data reduction was carried out with the LONG context in MIDAS using standard algorithm.

\section{Results}       
\label{sect:res}

Measured values of the equivalent widths and parameters estimates can be significantly affected by the spectral resolution of the data. The most illustrative example is the smearing of a P-Cyg profile at poor resolution, up to its disappearance. 
This lead to redistribution of the brightness in the line profile and a change in the value of its measured equivalent width. An accurate correction for the instrumental profile is often impossible without detailed modeling because the initial line profile is usually unknown and can be very complex (multicomponent). To use the grids with observations of low spectral resolution, we have smoothed the model spectra with Gaussian profile and constructed special versions of the diagrams. Figures \ref{fig:ew_contour_LBVhigh},~\ref{fig:ew_contour_ULXhigh} and \ref{fig:ew_contour_LBVlow},~\ref{fig:ew_contour_ULXlow} show an example of the EW diagrams in original resolution (without smoothing) and smoothed to a resolution of 5\,\AA, respectively. A careful comparison of Fig.\,\ref{fig:ew_contour_LBVhigh} with Fig.\,\ref{fig:ew_contour_LBVlow} (top right and bottom) and Fig.\,\ref{fig:ew_contour_ULXhigh} with Fig.\,\ref{fig:ew_contour_ULXlow} indicates that an impact of spectral resolution on the equivalent widths of the H$\alpha$ and He\,I $\lambda$5876 is minimal due to the weakness of the absorption components of these lines. More significant contribution of the absorption component to the EWs can be seen in diagrams for H$\beta$ and Fe\,II $\lambda$5169 (see Fig.\,\ref{fig:ew_contour_LBVhigh} and \ref{fig:ew_contour_LBVlow}, top left). The discrepancy in the temperature and mass-loss rate estimates between the high and low resolution versions of the H$\beta$--Fe\,II $\lambda$5169 diagrams is $\Delta T \approx 500\,$K, $\Delta \dot{M} \approx 0.05\,$dex for objects with temperatures $\gtrsim 19000\,$K and $\Delta T \approx 2000\,$K, $\Delta \dot{M} \approx 0.1\text{--}0.2\,$dex for objects with temperatures in the range $12000$--$16000\,$K.

We compared the stellar parameters of AG Car in hot state obtained from our equivalent width diagrams of the He\,II $\lambda$4686 and H$\alpha$ lines with the results of detailed modeling. Our estimates of the temperature $T_{*}\approx27000$\,K and the mass-loss rate $\dot{M}_{\text{cl}}\approx4.9\times10^{-5}\,M_{\odot}\text{yr}^{-1}$ are in good agreement with the values $T_{*}=26200$\,K and $\dot{M}_{\text{cl}}=6.3\times10^{-5}\,M_{\odot}\text{yr}^{-1}$ ($\log{L_{*}}=6.17$, $V_{\infty}=300\,\text{km s}^{-1}$, $f=0.1$) presented by \cite{Groh2009}.

Also, we compared the mass-loss rate and temperature estimates of P\,Cygni obtained from the H$\alpha$--Si\,II $\lambda$6371 diagram (Fig.\,\ref{fig:ew_contour_LBVlow}, top right) with accurate model parameters from the most recent work \cite{Rivet2019}. The difference between the model grid estimates $T_{*}\approx18500$\,K, $\dot{M}_{\text{cl}}\approx 4.4\times10^{-5}\,M_{\odot}\text{yr}^{-1}$ and the  accurate parameter values $T_{*}=18700$\,K, $\dot{M}_{\text{cl}}=4.0\times10^{-5}\,M_{\odot}\text{yr}^{-1}$ is $\lesssim 2\%$ for the temperature and $\approx 9\%$ for the mass-loss rate ($\log{L_{*}}=5.79$, $V_{\infty}=185\,\text{km s}^{-1}$, $f=0.5$).

Positions of LBVs Var\,A-1 and Var\,15 in the diagram H$\beta$--Fe\,II $\lambda$5169 (Fig.\,\ref{fig:ew_contour_LBVlow}, top left) are marked with triangles and rhombuses, respectively. Marker colors correspond to different years of the LBV observations: 2010 (black), 2013 (grey), 2015 (white). The diagram Fig.\,\ref{fig:ew_contour_LBVlow} (top left) shows that Var\,A-1 have moved to higher ionization state. In the case of Var\,A-1, the EW of the H$\beta$ emission component changes slightly $\approx 12\%$ between two states, however, a decrease in the EW of the Fe\,II and the H$\beta$ absorption component indicates a significant change in the ionization structure of the wind.

The Galactic LBV WS-1 have He\,II $\lambda$4686 line in the spectrum but in general this star shows a spectrum with Si\,II, N\,II and Fe\,III lines that correspond to temperatures $T_{*}<23000\,$K \cite{Kniazev2015}. The temperature estimate obtained from  He\,I $\lambda$5876--He\,II $\lambda$4686 diagram is $T_{*}\approx 27000\,$K. However, the temperature and mass-loss estimates obtained from H$\beta$--Si\,II $\lambda$6371 diagram are $T_{*}=15200$\,K, $\log{\dot{M}}=-4.72\,M_{\odot}\text{yr}^{-1}$ and $T_{*}=18800$\,K, $\log{\dot{M}}=-4.87\,M_{\odot}\text{yr}^{-1}$ (the counters of the line EWs intersect twice, Fig.\,\ref{fig:ew_contour_LBVlow}). The temperature estimates $T_{*}\approx 19000\text{--}22000$\,K looks more reliable for this object (black circle on Fig.\,\ref{fig:ew_contour_LBVlow}, top right) due to the presence of Si\,II, N\,II, Fe\,III lines in the spectrum. Probably, electron scattering make a significant contribution to the strength of the He\,II $\lambda$4686 line, which leads to an overestimation of the temperature determined from the $\lambda$5876--He\,II $\lambda$4686 diagram. This can be related to an extended photosphere or high turbulent velocity.

Blue and red spectra of UGC\,6456~ULX together with the best-fit model are presented in Fig.\,\ref{fig:ugc}. There is a small contribution of narrow lines of the surrounding nebula in the observed spectra, which could not be completely subtracted during extraction. The spectra have low signal-to-noise ratio due to the low brightness of the object (m$_V\approx20.3^m$). The bolometric luminosity at the distance to the galaxy of 4.54 Mpc \cite{Tully2013} equals to $\approx 2\times 10^6$~L$_{\odot}$ (reddening of A$_V = 0.2 \pm 0.1$). After the accurate modeling we increased the terminal velocity to 2100 km/s and $\beta$ to 1.35. The modeling yields the temperature and mass-loss rate of $T_{*}=31250$\,K and $\dot{M}=2.7\times 10^{-5}\,M_{\odot}\text{yr}^{-1}$ for filling-factor $f=0.3$, while the  temperature estimated from the grid is about $33000^{+2100}_{-750}$~K (see Fig.\,\ref{fig:ew_contour_ULXlow}). The mass-loss rate obtained from the grid and scaled to the observed luminosity and wind velocity using eq. (\ref{eq1}) is $7.6^{+2.0}_{-3.3}\times10^{-5}\,M_{\odot}\text{yr}^{-1}$ (the original value was $1.2^{+0.3}_{-0.5}\times 10^{-6}\,M_{\odot}\text{yr}^{-1}$). The large error of the mass-loss rate is caused by the uncertainty of the nebula contribution to H$_{\alpha}$ equivalent width. So, the parameters obtained from the detail modeling and from the grid are consistent within $3\,\sigma$ error for $T_{*}$ and within $2\,\sigma$ error for $\dot{M}$, although there is a large difference between $\dot{M}$ values.

Another three ULXs are located in the same area of the diagram as UGC\,6456~ULX. Their temperatures are in the range of 34000-36000~K, the mass-loss rates derived using the He\,II-H${\alpha}$ diagram are in the range of $(0.9 - 1.3) \times 10^{-6}\,M_{\odot}\text{yr}^{-1}$. For NGC\,4559~X-7 we additionally estimates its parameters using the He\,II-He\,I\,$\lambda5876$ diagram and found them consistent within 1\,$\sigma$ (T$_{*}$) and 3\,$\sigma$ errors ($\dot{M}$). To correct derived mass-loss rates for the observed bolometric luminosities of these ULXs, we calculated them using the V-band absolute magnitudes taken from \cite{Tao2011}, the reddening from \cite{Tao2011} and \cite{Vinokurov2018}, the temperatures obtained via our modeling and the wind velocities roughly estimated from the H$_{\alpha}$ line widths \cite{Fabrika2015}. After these corrections the outflow rates of each ULX become within the range $(1.1 - 2.8) \times 10^{-5}\,M_{\odot}\text{yr}^{-1}$.

\section{Discussion and conclusions}       
\label{sect:disc}

In this work we presented the grids of models of extended atmospheres implemented in a form of diagrams of equivalent widths of various emission lines observed in optical spectra of LBV-like stars and ultraluminous X-ray sources. The main idea of this work was to provide a tool for a simple and quick estimation of mass-loss rate and temperature in winds of these objects. We found that the wind parameters obtained from the grids and from the detailed modeling show a relatively good agreement. At least for LBV-like stars, as it was shown on the example of AG Car, the values of $\dot{M}$ and $T_{*}$ obtained by us are in agreement with the results by other authors. This allows us to consider the grids as a good alternative to complex and time-consuming calculations in those cases when precise determination of the wind parameters is not needed. In the future we plan to make the CMFGEN models of our grid public which will allow other scientists to use them as initial approximations in their calculations.

At present, there are still no reliable estimates of outflow rates of ultraluminous X-ray sources. Available estimates ($\sim 10^{-5}\text{--}10^{-4}$ M$_{\odot}$\,yr$^{-1}$) are either based on a comparison of these objects with the only known Galactic supercritical accretor SS433 \cite{Fabrika2015} or derived from X-ray luminosities of ULXs. Estimates of photosphere temperatures are also vary in a wide range (from a few of $10^4$ to $10^5$~K) depending on initial assumptions (for example, \cite{Fabrika2015,Tao2011,Tao2012}). The values of $\dot{M}$ and T obtained in this work do not contradict previous estimations.
  
The presence of the strong gas outflows in ULXs has been revealed by X-ray spectroscopic studies of the last few years. This studies discovered relatively weak blue shifted ($\sim 0.2\,c$) emission and absorption lines, which provide a direct evidence for the so-called "ultrafast outflows" (UFO, \cite{Pinto2016,Pinto2017}). For the ULXs, whose spectra were considered in this paper, the UFO was reliably found only in NGC\,5204 X-1; for Holmberg\,II X-1 there is a marginal detection \cite{Pinto2020}, and for the remaining two ULXs (NGC\,4559 X-7 and UGC\,6456 ULX) we did not find any UFO studies. Velocities of the UFO are $1.5 - 2$ orders of magnitude higher than the velocities that we obtained by measuring widths of emission lines and modeling the optical spectra of the ULXs. Such a difference indicate that the X-ray and optical spectra are formed in different areas of the wind. Indeed, 2D-RHD simulations of the outflow from the supercritical disk surface by various authors showed heterogeneity of the wind: a mildly-relativistic funnel outflow (non-collimated "jet") with gas temperatures of T$_{\text{gas}} \sim 10^{7.5}\text{--}10^{8}$ K is formed along the rotation axis, while colder dense slow gas (T$_{\text{gas}} \sim 10^{6.5}$ K, $\rho \sim 10^{-6}$ g\,cm$^{-3}$, $v< 0.01\,c$) is ejected from more distant areas of the disk in other directions (see for example, \cite{Kawashima2012}).

Also we have to note that the presence of the funnel in the ULX winds violates their spherical symmetry, which in some cases can make the CMFGEN models inapplicable. However, the currently available data do not show strong discrepancies between the observed and model spectra, and, apparently, if the influence of such an inhomogeneity on the observed spectra do exist, it is not significant.

\begin{acknowledgments}
The reported study was funded by RFBR  according to the research project N 18-32-20214.
A. K. is grateful to the Russian Foundation for Basic Research (grant N 19-02-00311).
S. F. is grateful to the Russian Foundation for Basic Research (grant N 19-02-00432).
Observations with the SAO RAS telescopes are supported by the Ministry of Science and Higher Education of the Russian Federation (including agreement No05.619.21.0016, project ID RFMEFI61919X0016).
\end{acknowledgments}

\bibliographystyle{AstroBull}
\bibliography{bibtexbase}

\end{document}

%% file: sao_cmd_author.tex
\def\saoname{Special Astrophysical Observatory,  Russian Academy of Sciences,
              Nizhnii Arkhyz, 369167 Russia}

\def\squareforqed{\hbox{\rlap{$\sqcap$}$\sqcup$}}

\def\sq{\ifmmode\squareforqed\else{\unskip\nobreak\hfil
\penalty50\hskip1em\null\nobreak\hfil\squareforqed
\parfillskip=0pt\finalhyphendemerits=0\endgraf}\fi}

\def\utw{\smash{\rlap{\lower5pt\hbox{$\sim$}}}}

\def\udtw{\smash{\rlap{\lower6pt\hbox{$\approx$}}}}

\def\diameter{{\ifmmode\mathchoice
{\ooalign{\hfil\hbox{$\displaystyle/$}\hfil\crcr
{\hbox{$\displaystyle\mathchar"20D$}}}}
{\ooalign{\hfil\hbox{$\textstyle/$}\hfil\crcr
{\hbox{$\textstyle\mathchar"20D$}}}}
{\ooalign{\hfil\hbox{$\scriptstyle/$}\hfil\crcr
{\hbox{$\scriptstyle\mathchar"20D$}}}}
{\ooalign{\hfil\hbox{$\scriptscriptstyle/$}\hfil\crcr
{\hbox{$\scriptscriptstyle\mathchar"20D$}}}}
\else{\ooalign{\hfil/\hfil\crcr\mathhexbox20D}}%
\fi}}

